\documentstyle[preprint,aps]{revtex}
\begin{document}
\preprint{ HU-SEFT-R-1995-16a}
\newcommand{\st}{\scriptstyle}
\newcommand{\sst}{\scriptscriptstyle}
\newcommand{\mco}{\multicolumn}
\newcommand{\epp}{\epsilon^{\prime}}
\newcommand{\vep}{\varepsilon}
\newcommand{\ra}{\rightarrow}
\newcommand{\ppg}{\pi^+\pi^-\gamma}
\newcommand{\vp}{{\bf p}}
\newcommand{\ko}{K^0}
\newcommand{\kb}{\overline {K^0}}
\newcommand{\al}{\alpha}
\newcommand{\ab}{\overline {\alpha}}
\def\be{\begin{equation}}
\def\ee{\end{equation}}
\def\bea{\begin{eqnarray}}
\def\eea{\end{eqnarray}}
\def\CPbar{\hbox{{\rm CP}\hskip-1.80em{/}}}%temp replacement due to no font
\def\ap#1#2#3   {{\em Ann. Phys. (NY)} {\bf#1} (#2) #3.}
\def\apj#1#2#3  {{\em Astrophys. J.} {\bf#1} (#2) #3.}
\def\apjl#1#2#3 {{\em Astrophys. J. Lett.} {\bf#1} (#2) #3.}
\def\app#1#2#3  {{\em Acta. Phys. Pol.} {\bf#1} (#2) #3.}
\def\ar#1#2#3   {{\em Ann. Rev. Nucl. Part. Sci.} {\bf#1} (#2) #3.}
\def\cpc#1#2#3  {{\em Computer Phys. Comm.} {\bf#1} (#2) #3.}
\def\err#1#2#3  {{\it Erratum} {\bf#1} (#2) #3.}
\def\ib#1#2#3   {{\it ibid.} {\bf#1} (#2) #3.}
\def\jmp#1#2#3  {{\em J. Math. Phys.} {\bf#1} (#2) #3.}
\def\ijmp#1#2#3 {{\em Int. J. Mod. Phys.} {\bf#1} (#2) #3.}
\def\jetp#1#2#3 {{\em JETP Lett.} {\bf#1} (#2) #3.}
\def\jpg#1#2#3  {{\em J. Phys. G.} {\bf#1} (#2) #3.}
\def\mpl#1#2#3  {{\em Mod. Phys. Lett.} {\bf#1} (#2) #3.}
\def\nat#1#2#3  {{\em Nature (London)} {\bf#1} (#2) #3.}
\def\nc#1#2#3   {{\em Nuovo Cim.} {\bf#1} (#2) #3.}
\def\nim#1#2#3  {{\em Nucl. Instr. Meth.} {\bf#1} (#2) #3.}
\def\np#1#2#3   {{\em Nucl. Phys.} {\bf#1} (#2) #3.}
\def\pcps#1#2#3 {{\em Proc. Cam. Phil. Soc.} {\bf#1} (#2) #3.}
\def\pl#1#2#3   {{\em Phys. Lett.} {\bf#1} (#2) #3.}
\def\prep#1#2#3 {{\em Phys. Rep.} {\bf#1} (#2) #3.}
\def\prev#1#2#3 {{\em Phys. Rev.} {\bf#1} (#2) #3.}
\def\prl#1#2#3  {{\em Phys. Rev. Lett.} {\bf#1} (#2) #3.}
\def\prs#1#2#3  {{\em Proc. Roy. Soc.} {\bf#1} (#2) #3.}
\def\ptp#1#2#3  {{\em Prog. Th. Phys.} {\bf#1} (#2) #3.}
\def\ps#1#2#3   {{\em Physica Scripta} {\bf#1} (#2) #3.}
\def\rmp#1#2#3  {{\em Rev. Mod. Phys.} {\bf#1} (#2) #3.}
\def\rpp#1#2#3  {{\em Rep. Prog. Phys.} {\bf#1} (#2) #3.}
\def\sjnp#1#2#3 {{\em Sov. J. Nucl. Phys.} {\bf#1} (#2) #3.}
\def\spj#1#2#3  {{\em Sov. Phys. JEPT} {\bf#1} (#2) #3.}
\def\spu#1#2#3  {{\em Sov. Phys.-Usp.} {\bf#1} (#2) #3.}
\def\zp#1#2#3   {{\em Zeit. Phys.} {\bf#1} (#2) #3.}
\def\zpp{$0^{++}$}
\def\fz{$f_0(980)$}
\def\az{$a_0(980)$}
\def\Kz{$K_0^*(1430)$}
\def\fzz{$f_0(1200)$}
\def\fzzz{$f_0(1200-1300)$}
\def\azz{$a_0(1450)$}
\def\ss{$ s\overline  s $}
\def\uu{$u\overline  u+d\overline  d$}
\def\qq{$q\overline  q$}
\def\KK{$K\overline  K$}
\def\sig{$\sigma$}
\def\lsim{\;\raise0.3ex\hbox{$<$\kern-0.75em\raise-1.1ex\hbox{$\sim$}}\;}
\def\gsim{\raise0.3ex\hbox{$>$\kern-0.75em\raise-1.1ex\hbox{$\sim$}}}
\setcounter{secnumdepth}{2}

\title{SUMMARY OF GLUONIUM95 AND HADRON95 CONFERENCES}
\baselineskip 0.15cm
\author{Nils A. T\"ornqvist }

\address
{University of Helsinki, SEFT, P.O. Box 9, Fin-00014 Helsinki, Finland}

\maketitle

\begin{abstract}\baselineskip .1cm
A short summary of two conferences on
the hadron and gluonium spectrum is given, with personal comments on
the status of  the best candidates for
gluonium or non-$q\overline  q$
states, such as $f_0(1370),\ f_0(1500),\ f_{0/2}(1720)$,  $f_J(2230)$,
 $\eta(1410),\ \eta(1460),$ and $f_1(1420)$.
\end{abstract}

\section{Introduction}

Since, at this conference\footnote{\baselineskip 0.1cm
{\it International Europhysics Conference on High Energy Physics},
Brussels, July 27th - August 2 1995. To be published in Conference proceedings
by World Scientific Publishing Company}, there is no plenary talk on hadron
spectroscopy, the organizers gave me the opportunity to
summarize two recent conferences, which were especially devoted to
this field.
This is thus a short personal summary of two conferences on
possible gluonium candidates and on the hadron spectrum held this summer.
The first was Gluonium'95 held in Propriano, Corsica, 30.6.-4.7. 1995,
and the second was Hadron'95 or the "6th International Conference on Hadron
Spectroscopy" held in Manchester 9.7-14.7 1995.
Of course, I cannot do justice to all interesting results reported at these
conferences. I shall, in fact, devote most of my time  on the new results
on non-$q\overline q$ candidates. For summaries of Hadron95 see
S.U. Chung's and M.R. Pennington's summaries
 in the Hadron95 proceedings\cite{proc}.

\section{Gluonium}

Gluonium or glueball states are the missing links in the standard model, which
predicts that these states should exist beyond any reasonable doubt.
Most people would agree that the search for these states is just as
important as the search for the top quark or the Higgs boson.
 If these gluonium states do not
exist, it would be a serious blow to our understanding of QCD. Thus
to find these states is one of the most important tasks for all
experimental groups in high energy physics. Unfortunately the
search has not yet been quite succesful. We have no "gold plated"
gluonium state, only a few good "candidates". One serious problem
in sorting out the experimental candidates is that we havn't had
a good enough model to treat broad and light $q\overline q$
 states, by which one
can distinguish these from gluonium. In addition, almost certainly there exists
 4-quark states, at least in the form of meson-meson
bound states. These mess up the meson spectrum, and for these our models are
even less developed.

At Propriano a rather informal meeting, devoted mainly to these
gluonium candidates,
was held with  33 registered participants. The small number
of participants allowed for a very loose organization. Thus there
 was no advance schedule of  speakers, the talks and discussions were usually
decided the same or the previous day, and all who wanted to contribute were
given time. This gave the meeting a special relaxed atmosphere where
ideas and results could be communicated in a spontaneous way.

We learned that there had been a remarkable advance in the results from
lattice gauge theory calculations. Although one is still far from having
very reliable results, the different groups now  agree that the lightest
glueball, which should be a flavour singlet $0^{++}$ state,
 should exist in the 1.5-1.75 GeV region, while the first tensor
or pseudoscalar  glueball
is expected somewhere near  2.15-2.45 GeV. The actually favoured numbers from
the
IBM group of D.  Weingarten et al.\cite{wein}
 is for $0^{++} 1740\pm71$ MeV and for $2^{++} \approx$2400
MeV, while UKQCD reports\cite{ukqc}
for the  $0^{++} 1550\pm50$ MeV and for the $2^{++} 2270\pm100$
 MeV. This is a substantial improvement in narrowing down the glueball
masses to rather small intervals. The improvement is mainly due to
the fact that todays computers allow for of the order of 30000 field
configurations on the lattice instead of previously
about 3000. Of course all this is
still in the quenched or valence approximation, i.e., whithout quark loops,
which is a very  drastic simplification of the actual situation.
Generally, from quantum mechanics, if one adds the new degrees
freedom due to the coupling to the multi-hadron continuum
states, then the gluonium masses should be shifted down in mass, since
the dominant part
of the hadron continuum is above the pure glue gluonium  mass.
However, this argument is not 100\% full proof, since presumably part of this
mixing with the continuum is phenomenologically already taken into account by
the fixing of scales in the pure glue lattice gauge theory calculations.

As to the magnitude of glueball widths one is usually rather
vague, but it is generally
believed that they should  be smaller or at most equal to normal
hadronic widths. A simple  argument for a smallish width is
based upon the fact that a pure glue
system must produce at least two $q\bar q$ pairs
in order to make the transition into
two normal hadrons, compared to only one pair for
a $q\overline q$ meson. Therefore, the gluonium width should roughly be of
the order of the geometric average of a normal meson width and a small
 OZI rule violating
width as, say $\phi\to \rho\pi$.
This would be important to elaborate upon,
 because one argument for why we havn't seen a gold plated
glueball state is that glueballs are extremely
broad. Then  we could not easily distinguish them from a smooth background.
 Weingarten\cite{wein} calculates a width of the scalar
gluonium to two pseudoscalars of $108\pm29$ MeV.
The naive prediction of flavour isotropic glueball
decay is disputed because of form factor effects, such that decays with
large phase space are suppressed. It was also argued  that
glue should be strongly connected to $\eta$'s and $\eta'$'s following
models of Gershtein\cite{gers} and Frere et al.\cite{frer}.

Two recent experimental glueball candidates were discussed at some
length. One was the Crystal barrel $f_0(1500)$\cite{crys},
which Amsler and Close\cite{clos} have argued is a strong candidate for a
glueball. Certainly, the mass is right if one believes the UKQCD
lattice calculation.
Also the fact that it is seen in the "gluon rich" channels is an
 argument in favour of a gluonium interpretation. Such channels are
radiative $J/\psi\to\gamma 4\pi$ decay
 and  central production or production by two pomerons.
The $f_0(1500)\ $ is produced both in radiative $J/\psi$ decay
(see the reanalysis of D. Bugg et al. discussed below) and in central
production by GAMS, (as the "$f_0(1590)$").
In addition, signals for the $f_0(1500-1590)$
appear to be prominent  in decay channels involving
$\eta$ and $\eta'$, which as discussed above favours gluonium.
However, the nearness of the important  $\rho\rho$ and $\omega\omega$
thresholds to the $f_0(1500)$ should make one seriously consider the
possibility that the $f_0(1500)$
may be a loosely bound
meson-meson deuteronlike bound state  or deuson\cite{deus}.

  The second glueball candidate is the
Beijing results on the narrow, $\approx$20 MeV, $f_J(2230)=
\xi (2230)$\cite{proc}, which was seen
first by Mark II in  $J/\psi\to \gamma K\overline  K$. Now BES sees it also in
$\gamma\pi\pi$ and  $\gamma p\overline  p$. The fact that the reduced $\pi\pi$
and $K\overline  K$ widths are nearly equal speaks for flavour isotropic decay,
and gluonium interpretation. Also the  production in the gluon-rich
$J/\psi$ radiative decay can be argued to favour a gluonium
interpretation. The $\xi(2230)$ could be the tensor glueball if $f_0(1500)$
is the scalar. But, the fact that it lies
precicely  at the $\Lambda\overline  \Lambda$ threshold makes
one a little suspicious
- maybe it could be a  $\Lambda\overline  \Lambda$ bound state.
The most serious non-$q\overline  q $ candidates are listed in table 1.

At Gluonium'95 there was one half day devoted to informal discussions of
how to interpret the  problematic mesons and meson candidates.
During this discussion one reached the remarkable
 agreement that the LEAR $f_0(1500)$
and the GAMS $f_0(1590)$ could be the same state,
 in spite of the fact that the experiments find different branching ratios.
In particular the $4\pi^0$ decay channel is found to be
large by the Crystal Barrel for the $f_0(1500)$,
while GAMS found large branching ratios $f_0(1590)\to \eta\eta$ and
$\eta\eta'$, while $f_0(1590)\to 4\pi^0$ was small.

\section{Light mesons}

\subsection{Light scalars}

At the Manchester Hadron95 conference there was much interest in the
controversial scalar meson sector. The lightest scalar nonet were discussed
in theoretical models by M. Scadron\cite{proc} and myself\cite{scal}.
Also new interesting results are now emerging as
many experimental papers now have begun
to include the sigma meson in their analysis
of multipion channels. Here the $\sigma$ is a very broad
structure peaking around 600-900 MeV and which is analytically
described by the $\pi\pi$ phase shifts.

There was an interesting reanalysis by D. Bugg et al.\cite{proc} of Mark III
data on $J/\psi\to \gamma\pi^+\pi^-\pi^+\pi^-$. This was previously
analysed assuming $\rho\rho$ dominance in the four pion system. Then one
found the puzzling result that pseudoscalar resonances dominate
the $\rho\rho$ mass spectrum, whose masses
 did not agree with any previously seen states.
Now by including $\sigma\sigma$ in the 4 pion system the authors find
much more reasonable results. They find a superior fit with
I=0  resonances at 1505, 1750, and 2104 MeV, as well as the conventional
resonances $f_2(1275)$, $f_2(1640)$ and $\eta(1440)$. The I=0 resonances
decay predominantly to $\sigma\sigma$. It is natural to identify
their I=0 resonance at 1505 with the LEAR glueball candidate $f_0(1500)$.

There was much discussion on the mass of the $4\pi$ resonance listed
in the 1994 tables\cite{pdg} under the entry
$f_0(1370)$. In a previous analysis by Gaspero\cite{gasp},
in which he reanalysed old Rome-Syracuse
bubble chamber data with 4 charged pions,
including  for the first time the $\sigma\sigma$ intermediate channel,
he found a $1386\pm30$ MeV mass value.
This was surprising, since the 4 pion mass distribution
peaks clearly at a higher mass, near 1500 MeV, and the analysis needed only
one resonance and no large interfering background, which could shift the peak.
Also the Crystal barrel reported
a mass  1374$\pm38$ MeV in $p\overline  p\to (\pi^+\pi^-2\pi^0)\pi^0$, and
OBELIX a mass of $1345\pm12$ MeV in
$\overline  n p\to (2\pi^+2\pi^-)\pi^+$. All these analyses
had a very broad width of almost 400 MeV.
I have been skeptical for long about this low mass value for the $f_0(1370)$,
suspecting that there was a mistake in flux or phase space factors in
the analyses.

Therefore, I was happy to hear that
now S. Resag\cite{proc}, also from Crystal Barrel finds in a careful
analysis of $p\overline  p\to 5\pi^0$ data with large statistics no
resonance at 1370 MeV
in the  $4\pi^0$ system, but instead
 a resonance at the mass of 1500$\pm10$ MeV and with
a width of $185\pm 20$ MeV.
He  also includes the decay mode  $f_0\to\sigma\sigma\to 4\pi^0$
 and finds that the 1500 decays mainly into $\sigma\sigma$.
 Now, if $f_0(1370)\to \sigma\sigma$ would be
present in the charged pions, as the previous
three analyses found, it must
by isospin be even more clearly be seen in 4 neutral pions,
where there is no $\rho\rho$ background.
Thus the "$f_0(1370)$" must, if it exists,
show up in Resag's  analysis, but he finds only the $f_0(1500)$.
In another paper on the same question of the mass of the $f_0(1370)$,
Achasov and Shestakov\cite{proc} conclude that the true mass of this $4\pi$
resonance cannot be at 1370 MeV, but must lie above 1500 MeV.
If this is so, it seems clear that the mass of the $f_0(1370)$ (at least
in the $4\pi$ channel) is much too low  in the 1994 PDG tables. If it
is instead  around 1500 MeV, then the $f_0(1370)$ in the 4$\pi$ mode
and $f_0(1500)$
are likely to be the same resonance, while the entries under $f_0(1370)$
of the PDG1994\cite{pdg} in the two pseudoscalar mode are likely
to belong to the $f_0(1300)$ . Hopefully the
situation will be cleared up till
the next conference, since the $f_0(1500)$
is an important non-$q\overline  q$  and
glueball candidate, while $f_0(1370)$ is an extra state, which does not have
an obvious place in the meson spectrum.

\subsection{ The $\eta(1410),\ \eta(1460)$ and $ f_1(1420)$}

Many new results were reported on these controversial mesons, and we
are gradually getting a much better picture of what is really observed.

In central production of the $K_S K^{\pm}\pi^{\mp}$ system in $pp$
reactions with a $LH_2$ target and a  800 GeV/c beam
the E690 experiment clearly sees the  $f_1(1420)$, which predominantly
decays into $K^*\overline  K+ c.c.$. They have a very clear peak at 1420 MeV.
A Dalitz plot analysis shows an interference pattern of the two $K^*$
bands, which very convincingly
 is consistent only with a spin parity assignment of $1^{++}$.
Any background of $0^{-+},\ 1^{+-}$ or $1^{-+}$ is small. This spin parity
is also supported by the fact that the $f_1(1420)$ is also seen in
$\gamma\gamma^*$ production.
Thus in central production one produces the $f_1(1420)$, which by now
is a very well established resonance, but no pseudoscalar
 $\eta(1400-1460)$ is observed in central production. The $f_1(1420)$ is very
likely a non-$q\overline  q$
resonance (see table 1), since the $f_1(1520)$ already completes the $1^{++}$
nonet as the favoured $s\overline  s$ state.     The $f_1(1520)$
must also be considered as a well established resonance, since it is clearly
seen
in four experiments\cite{pdg} and in three reactions:
$\pi^- p\to (K^+\overline  K^0 \pi^-) n$,
$K^-p\to (K_SK^{\pm}\pi^{\mp})\Lambda$, and
$\gamma\gamma^*\to \pi^+\pi^-\pi^0\pi^0$.
On the other hand only $f_1(1420)$, not $f_1(1520)$, is seen in central
production, which clearly shows the very different nature of these two axial
resonances.

The extra state, the $f_1(1420)$, is generally not believed to be a
 glueball, since its mass is too low. Possibily it could be a hybrid state, but
a more  likely situation is that it is a
4 quark state, probably in the form of
(a virtually bound) $K\overline  K^*$ system.

If so, it shows that multiquark states can be produced preferably
in the "gluon rich"
invironment of central production or radiative $J/\psi$ decay.
 This should be a warning to those who
want to use such environments as a good place to look for glueballs.

In $\overline N N $ annihilation studied
by the Crystal Barrel and OBELIX at LEAR
collaborations one does not see a strong $f_1(1420)$
but instead the "iota" pseudoscalars in the $\pi\pi\eta$ and $K\overline  K\pi$
systems. There is now mounting evidence, which show that the "iota" peak
is actually two resonances, one low mass decaying mainly into $a_0(980)\pi\to
\eta\pi\pi$
+ $(K\overline  K)_{S-wave}\pi$,
and a heavier one decaying into mainly $K\overline  K^*$.

In the $\eta\pi\pi$ decay mode the Crystal barrel sees an $\eta$ resonance
at $1409\pm3$ MeV with a width of $86\pm10$ MeV. The  final state $\eta\pi\pi$
is reached via two intermediate states $\sigma \eta$ and $a_0(980)\pi$.
The $a_0(980)$ then decays both into $K \overline K$ and $\eta\pi$, such that
 the ratio $K\overline K/\eta \pi$ is 1.1$\pm 0.3$, when one integrates over
the $a_0$ peak.

The OBELIX collaboration has about 4000 $p\overline  p\to \pi\pi (K\overline
K\pi)$
events from a sample of 18 million annihilations at rest. They see
two peaks one lower (m=1415$\pm 2$ MeV , $\Gamma =59\pm4$ MeV) decaying
into $(K\overline  K)_{S-wave}\pi$ and a heavier one (m$=1460\pm10$ MeV,
$\Gamma =
100\pm10$ MeV), which decays into $K\overline  K^*+c.c$.

Certainly two pseudoscalars in the 1400-1460 MeV region is at least
one too many. One of the peaks can be the $s\overline  s$ partner of the
$\eta(1295)$ completing the $2^1S_0$ $q\overline  q$ nonet, but the second
one is a non-$q\overline q$ candidate. The mass is believed to be
too low for beeing a good gluonium candidate.
Thus it is perhaps more likely a 4 quark, $K\overline  K^*$ or possibly a
    hybrid state.

\section{Heavy hadrons}

There were reports from LEP and the Delphi collaboration by M.
Feindt\cite{proc}
with updates on $B$,$B_s$ and $\Lambda_b$ masses and lifetimes. Also better
determination of the $B^*$ mass were reported. More
interesting was  the first discoveries of
the orbitally excited $b\overline q\ $ P-wave
states, generically called $B^{**}$ and $B^{**}_s$. These would belong to the
$0^{++},1^{++},1^{+-}$ and $2^{++}$ multiplets of the normal quark model. They
are expected to lie very close to each others and should be overlapping, since
their widths should be larger than their mass splittings,
and the two axial states
should mix like already the $K_{1A}$ and $K_{1B}$ states do in the light
sector. Heavy quark
symmetry is here more useful than conventional flavour symmetry.
The mass of the $B^{**}$ meson is found to be $5712\pm11$ MeV by OPAL,
$5732\pm5\pm20$ MeV by DELPHI and $5734\pm3\pm18$ MeV by ALEPH.
The width is of the order 100 MeV. For the $B^{**}_s$ OPAL reports
$5884\pm15$ MeV and a width of $47\pm22$ MeV,  and the preliminary
results from DELPHI on the $B^{**}_s$ are similar.

{}From CLEO there were also interesting new results on the spectroscopy
of charmed states reported by J.
Bartelt\cite{proc}.  Perhaps the most interesting result
is the first observation of
the isospin violating $D_s^*\to D_s\pi^0$ decay, which can provide a good
measurement of the $\pi^0 -\eta$ mixing. Also, this decay mode
gives a new very accurate measurement of the $D^*_s - D_s$ mass splitting
of $143.76\pm 0.39\pm 0.40$ MeV.

For more details on heavy
meson spectroscopy one should of course consult the original papers.

\section {Panel discussion}

 At the hadron95 conference there was a special session with a round table
 discussion on general problems of the future for hadron spectroscopy.
Here people could express their views on what experiments should be
performed, which models should be studied, how many glueballs should be found,
which faclities are needed etc.

The scheduled shutdown of LEAR at the end of 1996 was very much
regretted.  Collegues working in other fields
of particle physics do not today seem to consider hadron spectroscopy as
important enough compared to, say,  the search for the top or the Higgs boson.
However, the elusive gluonium states are indeed more fundamental to the
understanding of the non-Abelian nature of QCD than anything else,
including the top. For spontaneous symmetry breaking and mass generation
the sigma  meson (which we still really don't know exactly where it is)
 is equally important for the masses of the nucleon, constituent quarks
and hadrons in general,
as the Higgs boson is for the masses of current quarks, leptons, W and Z.
Confinement and nonperturbative aspects of QCD are still
not understood and hadron spectroscopy is the crucial experimental input
for their theoretical understanding.

The many exciting results presented at these two conferences
from Beijing, BNL CERN, Fermilab, KEK, Serpukhov etc.
on glueball candidates as well as on
many other mesons show that the field is still very much alive, and that many
important new discoveries certainly lie ahead.

\begin{table}\begin{center}\caption{Main non-$q\overline  q$ candidates.}
\vspace{0.5cm}
\begin{tabular}{|l|l|} \hline\hline
$f_0(1500-1590)$ & Gluonium, $\rho\rho,\ \omega\omega$, or  4q state? \\
$f_{0/2}(1720)$ & Gluonium, $K^*\overline  K^*$, or  4q state? \\
$f_J(2230)$ & Gluonium, $\Lambda\overline  \Lambda$, or  4q state? \\
$\eta(1410),\eta(1460)$ & Gluonium, $K\overline  K^*$, or  4q state? \\
$f_1(1420)$ & Gluonium,    $K\overline  K^*$, or  4q state? \\
\hline %\cline{1-2}
\end{tabular}
\end{center}
\end{table}
\end{document}